\documentclass[12pt]{article}
\usepackage[top=1in, bottom=1in, left=1in, right=1in]{geometry}
\usepackage[english]{babel}
\usepackage{atbegshi,cite}
\usepackage{amsmath,amssymb,amsbsy,amstext, amsthm, simplewick}
\usepackage{hyperref}
\usepackage{graphicx}
\usepackage{amsfonts}
\usepackage[small]{caption}
\usepackage{upgreek}
\usepackage[titletoc]{appendix}
\usepackage{setspace}
\usepackage{color}

\newcommand{\newc}{\newcommand}
\newc{\fpi}{f_{\pi}}
\newc{\etap}{\eta^{\prime}}
\newc{\llll}{\langle\lambda\lambda\rangle}
\newc{\FFd}{F^a\tilde F^a}
\newc{\qbar}{{\overline q}}
\newc{\TR}{{\rm Tr}}
\newc{\Kahler}{K\"ahler }
\newc{\Zbb}{{\mathbb Z}}
\newc{\Rt}{{\mathbb R}^3}
\newc{\Rf}{{\mathbb R}^4}
\newc{\So}{{\mathbb S}^1}
\newc{\zt}{{\mathbb Z}_2}
\newc{\RtSo}{{\mathbb R}^3\times{\mathbb S}^1}
\newc{\scriminus}{{\cal I}^-}
\newc{\scriplus}{{\cal I}^+}
\newc{\mpl}{M_p}
\newc{\Ricci}{\mathcal{R}}
\newc{\bv}{\phi}
\newc{\calU}{{\cal U}}
\newc{\calK}{K}
\newc{\calUi}{{\cal U}^{-1}}
\newc{\calG}{{\cal G}}
\newc{\calO}{{\cal O}}
\newc{\calQ}{{\cal Q}}
\newc{\calOb}{{\cal O}^\dagger}
\newc{\hphi}{{\hat\phi}}

\theoremstyle{plain}
\theoremstyle{plain} 
\theoremstyle{plain} 
\theoremstyle{plain}
\theoremstyle{plain}
\theoremstyle{plain}

\renewcommand{\title}[1]{{\Large\bf\flushleft{#1}}\vspace*{3ex}\\}
\renewcommand{\author}[2]{{\noindent\hspace*{2.5em}\large#1}
                     \footnote{Electronic mail: $\mathtt{#2}$}\\}

\begin{document}
\begin{titlepage}
\begin{flushright}
{\large 
~\\
}
\end{flushright}

\vskip 2.2cm

\begin{center}

{\large \bf Transplanckian Censorship \\ and the Local Swampland Distance Conjecture }

\vskip 1.4cm

{ Patrick Draper$^{(a)}$ and Szilard Farkas}
\\
\vskip 1cm
{$^{(a)}$ Department of Physics, University of Illinois, Urbana, IL 61801}\\
\vspace{0.3cm}
\vskip 4pt

\vskip 1.5cm

\begin{abstract}

The swampland distance conjecture (SDC) addresses the ability of effective field theory to describe distant points in moduli space. It is natural to ask whether there is a local version of the SDC: is it possible to construct local excitations in an EFT that sample extreme regions of moduli space? In many cases such excitations exhibit horizons or instabilities, suggesting that there are bounds on the size and structure of field excitations that can be achieved in  EFT. 
Static bubbles in ordinary Kaluza-Klein theory provide a simple class of examples: the KK radius goes to zero on a smooth surface, locally probing an infinite distance point, and the bubbles are classically unstable against radial perturbations. However, it is also possible to stabilize KK bubbles at the classical level by adding flux. We study the impact of imposing the Weak Gravity Conjecture (WGC) on these solutions, finding that a rapid pair production instability arises in the presence of charged matter with $q/m\gtrsim 1$. We also analyze 4d electrically charged dilatonic black holes. Small curvature at the horizon imposes a bound $\log(M_{BH})\gtrsim |\Delta\phi|$, independent of the WGC, and the bound can be strengthened if the particle satisfying the WGC is sufficiently light. We conjecture that quantum gravity in asymptotically flat space requires a general bound on large localized moduli space excursions of the form $ |\Delta\phi|\lesssim |\log(R\Lambda)|$, where $R$ is the size of the minimal region enclosing the excitation and $\Lambda^{-1}$ is the short-distance cutoff on local EFT. The bound is qualitatively saturated by the dilatonic black holes and Kaluza-Klein monopoles.

\end{abstract}

\end{center}

\vskip 1.0 cm

\end{titlepage}
\setcounter{footnote}{0} 
\setcounter{page}{1}
\setcounter{section}{0} \setcounter{subsection}{0}
\setcounter{subsubsection}{0}
\setcounter{figure}{0}


\section{Introduction}
It is of interest to try to determine ways in which low-energy physics is constrained by consistent embedding in a quantum theory of gravity.
A number of conjectures have focused on the properties of moduli spaces.

For example, the swampland distance conjecture (SDC)~\cite{swampland2} states that homogeneous motion over large distances in any large moduli space results in a tower of exponentially light states descending below the cutoff of the initial EFT. A simple example of a gravitational theory with a large moduli space is ordinary Kaluza-Klein (KK) theory. The energy of the KK spacetime ${\cal R}^{D,1}\times S^1$ does not depend on the size $R$ of the circle, and the invariant distance between two points in the moduli space is $\int dR/R$, which diverges logarithmically as the circle size goes to zero or infinity. If one changes the asymptotic value of the modulus in this theory, a tower of states -- either KK states or wound string states -- becomes light.

These ideas are conceptually clear, and there has been considerable recent investigation of the SDC (see, for example,~\cite{Klaewer:2016kiy,Blumenhagen:2017cxt,Palti:2017elp,Hebecker:2017lxm,Grimm:2018ohb,Heidenreich:2018kpg,Hebecker:2018vxz,Scalisi:2018eaz,Palti:2019pca,Lust:2019zwm}). However, since a given EFT corresponds to fixed asymptotic values of the moduli, it is natural to pose a complementary question: is there a local version of the SDC? In other words, is there any limitation on localized excitations that sample distant regions of moduli space? Such obstructions might arise in a different way than the appearance of a tower of light states. 

In fact, a number of other rather disparate classical and semiclassical examples of this ``transplanckian censorship" phenomenon are known~\cite{tbanks, ArkaniHamed:2007js, nicolis, draperetal, Draper:2018lyw}.\footnote{This notion of transplanckian censorship is distinct from the recent conjectures in~\cite{Bedroya:2019snp,Bedroya:2019tba}.}  For example, in a 4d massless scalar field theory minimally coupled to gravity, static, spherically symmetric excursions of the scalar in regions of subplanckian curvature are bounded by $\calO(1)$ in Planck units~\cite{nicolis}. However, this theory can also be realized as the dimensional reduction of the 5d KK theory. In the KK theory there are  solutions known as KK bubbles that sample all the way to $R=0$ in a local region of low 5d curvature. KK bubbles are thus a concrete example of a localized excitation sampling an infinite distance in moduli space, and it is of interest to examine their properties in more detail.

Informally, KK bubbles describe spherical holes  of size $\rho_0$ in asymptotically KK space. 
Expanding  bubbles can nucleate nonperturbatively~\cite{BON}, and the description of this process as tunneling under an energy barrier was elucidated in~\cite{brillhorowitz}. One might already take Witten's bubble of nothing as an indication of the inconsistency of the theory. However, the lifetime of the ordinary KK vacuum can be exponentially long, and there are other static bubble excitations that exhibit more dramatic behaviors. Static ``Schwarzschild" bubble solutions were first found in~\cite{sorkin,GP}, along with a larger family of static ``Kerr" solutions. Near the wall of a KK bubble, the circle radius $R$ goes to zero, smoothly truncating the spacetime the physical radius $\rho_0$. The geometries therefore have the interesting property that they sample points separated by an infinite proper distance in moduli space in well-localized, low-curvature regions of physical space. From the perspective of dimensional reduction, the KK scalar diverges on the surface of the bubbles.  These solutions are thus a natural laboratory for the questions raised above.\footnote{Casimir energies lift the moduli space in nonsupersymmetric KK theories.  As usual we assume that the classical bubble solutions provide useful approximations to solutions in theories with moduli stabilized by additional fluxes or other objects. Ref.~\cite{Dine:2004uw}, for example, found that neutral bubble solutions persist after adding simple stabilizing potentials.}

It turns out that all of these static bubbles are classically unstable. The  instability of the static, asymptotically flat Schwarzschild bubble was demonstrated in~\cite{grossperryyaffe} and given a mechanical interpretation in~\cite{brillhorowitz}: this bubble sits at the top of the potential ``hill" under which Witten's bubble mediates tunneling.  (The static bubble is therefore also responsible for topology change at high temperature~\cite{brownthermal}, analogous to a sphaleron in gauge theories.)  Similarly, the asymptotically flat Kerr bubbles were shown to be unstable in~\cite{Draper:2019zbb}, with an equivalent relationship to a known tunneling process~\cite{dowkeretal}. It was suggested in~\cite{Draper:2019zbb} that the classically instabilities of the Schwarzschild and Kerr bubbles should be thought of as a pathology of the type described above: distant points in moduli space are ``hidden" behind an instability. 

It is also known, however, that KK bubbles can be perturbatively stabilized by embedding them in spacetimes with different asymptotics, or, in  asymptotically flat space, by wrapping them in flux. In the latter case, explicit examples of bubble geometries stabilized by 3-form flux were found in~\cite{Gibbons:1994vm,Horowitz:2005vp}. These spacetimes do not appear to be particularly theoretically exotic, and so it is curious that they do not seem to exhibit horizons or instabilities.

In a  different context, it has recently been shown in Refs.~\cite{Crisford:2017zpi,Crisford:2017gsb,Horowitz:2019eum} that potential counterexamples to {\emph{cosmic}} censorship can be avoided by imposing the weak gravity conjecture (WGC)~\cite{wgc}. In short, the proposed counterexamples involve electromagnetic fields, and when charged scalar fields satisfying $q/m>1$ are added,  the solutions are unstable against scalar perturbations. (Scalar fields are used to facilitate a classical analysis; fermions are expected to perform a similar function, but a more complicated treatment is required.)

We will apply the idea of~\cite{Crisford:2017zpi,Crisford:2017gsb,Horowitz:2019eum} to the perturbatively-stable charged KK bubble spacetimes of~\cite{Horowitz:2005vp} and argue that a new instability arises in the presence of charged matter satisfying the WGC. Charged objects are wound strings, and one can screen some of the bubble's charge by throwing oppositely charged strings into it. For sufficiently large $q/m$, we might expect that the vacuum will become unstable against rapid Schwinger production near the bubble wall. We study this question with a toy model in the dimensionally reduced theory, where the lowest wound string modes are represented by a massive charged scalar field coupled to ordinary electromagnetic flux. We show that in this model the negatively-charged ground state energy drops below $-m$ for $q/m\gtrsim 1$, signaling an instability against pair creation, and we argue that  discharge rate is typically much faster than the tunneling rate to larger expanding bubbles. This  suggests that the WGC can play a similar role in the censorship of infinite localized field excursions. 

Another interesting class of geometries is provided by charged black holes with large moduli variations outside the horizon. We estimate the discharge rate of 4d charged dilatonic black holes of~\cite{Garfinkle:1990qj}. In these geometries, the size of the dilaton excursion from infinity to the horizon is controlled by the charge of the black hole, diverging in the extremal limit. Low curvature in the region of the large excursion requires $|\Delta\phi|\lesssim \log(M)$. For sub-extremal black holes the excursion is finite, and we find that the discharge rate is fast if the WGC  is satisfied by a light particle of mass $m\ll1/M$. For sufficiently large black holes $M\gtrsim {\rm max}(e^{|\Delta\phi|},1/m)$ the rate is slow.

We conclude with a loose conjecture: quantum gravity in asymptotically flat space requires a general bound on large localized moduli space excursions of the form $ |\Delta\phi|\lesssim |\log(R\Lambda)|$, where $R$ is the size of the minimal region enclosing the excitation and $\Lambda^{-1}$ is a short-distance cutoff. Both neutral and charged KK bubbles have finite $R$ and infinite excursions, but are strongly unstable. Dilatonic black holes in a controlled EFT also satisfy the bound. KK monopoles provide another example: they are stable and sample an infinite distance in moduli space, but only at a single point, so the visible excursion is limited by the short-distance  cutoff.

\section{Charged KK Bubbles}

We begin by discussing a representative class of Kaluza-Klein bubbles perturbatively stabilized by flux. We then add matter of mass $m$ and charge $q$ to the system and demonstrate the existence of a rapid pair-production instability for $q/m\gtrsim 1$.

\subsection{Classical Solutions}
A number of static charged bubble solutions were obtained in~\cite{Gibbons:1994vm,Horowitz:2005vp}. In~\cite{Horowitz:2005vp}, a 6D bubble stabilized by electric and magnetic 3-form flux was constructed from a family of 5D zero-momentum initial data characterizing bubbles of different sizes. This method is particularly convenient for assessing bubble stability against radial perturbations. We review this construction here, simplifying to case of purely electric 3-form flux. 

A family of five-dimensional spatial metrics is given by
\begin{align}
ds^2_{spatial}=U(\rho)d\chi^2+\frac{d\rho^2}{U(\rho) h(\rho)}+\rho^2d\Omega_3
\label{eq:spatial}
\end{align}
with
\begin{align}
U(\rho)\equiv 1-\frac{\rho_0^2}{\rho^2}.
\label{eq:U}
\end{align}
The function $h$ will be determined by the Hamiltonian constraint, and the periodicity of the KK circle at infinity $\chi\sim\chi+L$ will be determined by smoothness of the metric at $\rho=\rho_0$. 

We now add electric 3-form flux to the bubbles, $C=\frac{Q_0}{2\pi^2}(\star \epsilon_3)$, where $\epsilon_3$ is the volume element of the spatial $S^3$. Concretely, the field strength is
\begin{align}
C_{\rho t\chi}=\frac{NQ_0}{2\pi^2 \sqrt{h}\rho^3}\;,
\label{eq:cN}
\end{align}
where $N$ is the lapse function. The Hamiltonian constraint is then\footnote{The matter Lagrangian is normalized as $-\frac{1}{12}\int d^6x\, \sqrt{-g_6}C^2$.}
\begin{align}
{}^5R=\frac{Q^2}{\rho^6}\;,
\end{align}
where $Q\equiv Q_0/(2\pi^2M_6^2)$ and $M_6=(8\pi G)^{-1/4}$ is the 6D Planck scale. We find
\begin{align}
h(\rho)\equiv1+\frac{b}{3\rho^2-2\rho_0^2}-\frac{Q^2}{4\rho_0^2\rho^2}
\label{eq:h}
\end{align}
where $b$ is an arbitrary constant.

To make the geometry smooth everywhere, we impose periodicity $\chi\sim\chi+L$ on the KK circle, where
\begin{align}
L=\frac{2\pi \rho_0}{\left(1+\frac{b}{\rho_0^2}-\frac{Q^2}{4\rho_0^4}\right)^{1/2}}\;.
\label{eq:b}
\end{align}
With this periodicity, space ends on a smooth cap at $\rho=\rho_0$. 

\begin{figure}[t!]
\begin{center}
\includegraphics[width=0.5\linewidth]{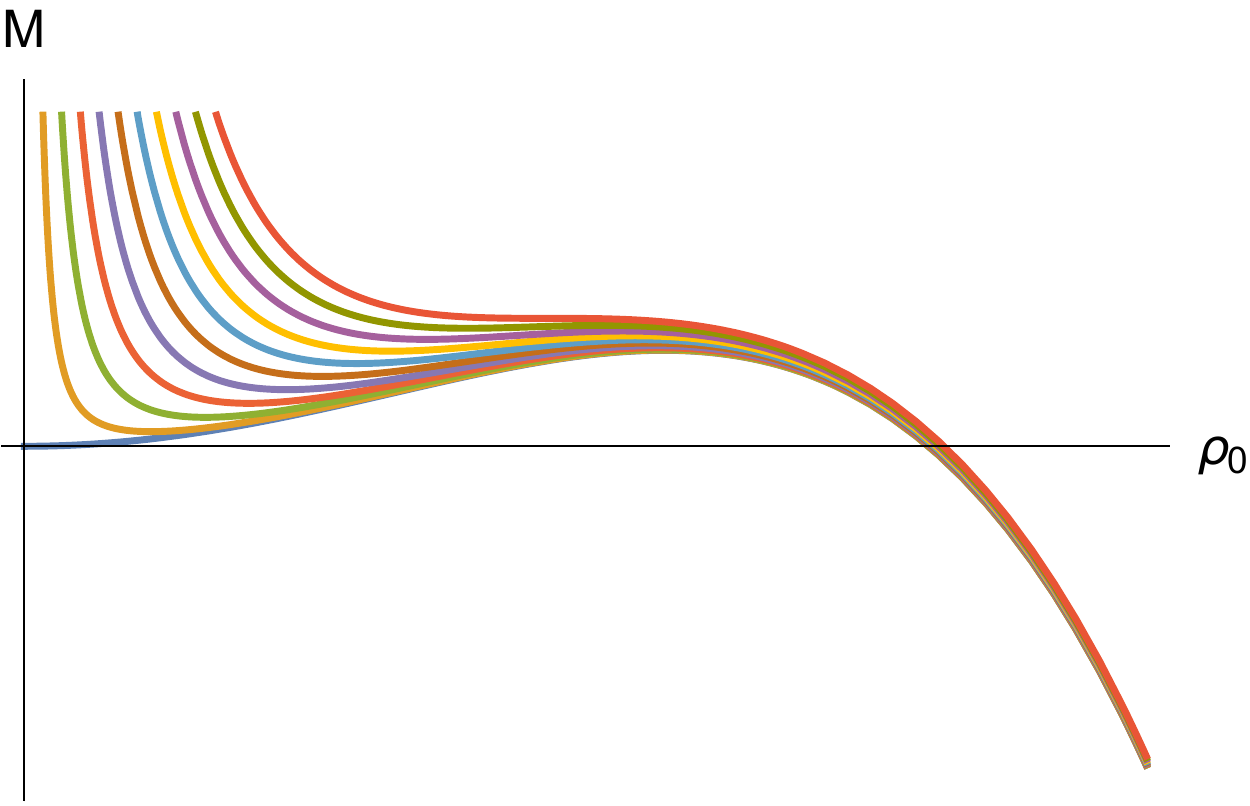}
\caption{ The energy (Eq.~(\ref{eq:Mbubble})) of a one-parameter family of initial data labeled by bubble radius  $\rho_0$. From bottom to top, curves correspond to $Q=0,\dots,Q=Q_{max}$. The most stable bubbles lie at small $Q$; for $Q>Q_{\max}$, no stable solution exists.
} 
\label{fig:Mrho0}
\end{center}
\end{figure} 

Thus far we have a family of charged bubble initial data. For a given charge $Q$ and asymptotic circle size $L$, it is a one-parameter family of bubbles labeled by the radius $\rho_0$. The energy of the family is
\begin{align}
M=\pi^2 LM_6^4\left(\frac{Q^2}{2\rho_0^2}+2\rho_0^2-\frac{4\pi^2\rho_0^4}{L^2}\right)\;.
\label{eq:Mbubble}
\end{align}
There is a stable local minimum of the energy $M(\rho_0)$ for all $Q$, $L$ such that
\begin{align}
Q<Q_{max}=\frac{L^2}{3\pi^2\sqrt{3}}\;.
\label{eq:stab}
\end{align}
There is also an unstable maximum at larger $\rho_0$. For $Q>Q_{max}$, there are no stationary points.

In Fig.~\ref{fig:Mrho0} we illustrate the mass of the initial data as a function of bubble radius $\rho_0$ for $0\leq Q\leq Q_{max}$. Clearly even the perturbatively stable bubble at the local minimum is unstable against tunneling to larger radii. At $Q= Q_{max}$, the barrier disappears completely. On the other hand, for small $Q$, the barrier grows, and the rate for this transition may become extremely slow. In the next section we will show that a new, rapid instability arises in this limit. 

In general the local minimum is a root of a cubic, equivalent to $b=0$ in Eq.~(\ref{eq:b}). For small $Q$ it simplifies to 
\begin{align}
\rho_0^2\approx Q/2.
\end{align}
It is straightforward to verify that the local minimum is a static solution to the full Einstein equations with spacetime metric
\begin{align}
ds^2=-h(\rho)dt^2+U(\rho)d\chi^2+\frac{d\rho^2}{U(\rho) h(\rho)}+\rho^2d\Omega_3\;
\label{eq:dsfull}
\end{align}
with $U$ and $h$ given by Eqs.~(\ref{eq:U}),~(\ref{eq:h}) and $b=0$. The mass of this static bubble is 
\begin{align}
M_{min}=\pi^2 LM_6^4\left(\frac{3Q^2}{4\rho_0^2}+\rho_0^2\right),
\end{align}
where $L$ is given by Eq.~(\ref{eq:b}) evaluated at $b=0$. The field strength~(\ref{eq:cN}) surrounding the static bubble simplifies to
\begin{align}
C_{\rho t\chi}=\frac{Q_0}{2\pi^2 \rho^3}\;.
\label{eq:cN}
\end{align}

The point in moduli space where the size $R$ of the KK circle vanishes lies an infinite proper distance $\int dR/R$ away from any point of finite circle size. $R=0$ is sampled locally on the wall of KK bubbles, since $V\rightarrow 0$ as $\rho\rightarrow\rho_0$, while $R=L$ at spatial infinity.

Typically, static neutral bubbles of nothing in asymptotically flat space have a single unstable mode, corresponding to perturbations of the bubble radius. The solution~(\ref{eq:dsfull}), lying at a local minimum of the energy, is perturbatively stabilized by the flux. It disappears for $Q=0$, leaving only the perturbatively unstable point corresponding to an ordinary neutral static bubble. We also see from the Hamiltonian constraint that for small $Q/L^2$, curvatures near the bubble are of order $1/Q$. Therefore there is a minimum $Q$, controlled by the cutoff, for which we can study this geometry classically. 

\subsection{Adding Charged Matter}
We would like to study the stability of the flux-stabilized bubble against the introduction of probe charges. In the 6D description, charged objects are wound strings. To simplify the analysis, we consider only the lowest states of a string with winding number one and zero KK excitations. Formally, we can dimensionally reduce over the  circle to obtain a 5D geometry with ordinary electromagnetic flux, and we introduce a massive  scalar particle with charge $q$ to represent the wound string. Near the bubble wall the scalar mass decreases with the radion.

In this toy model we  can study the single-particle ground states of positive and negative charge as a function of $q$. For $|q|/m\gtrsim 1$, the electrostatic potential energy of a negatively charged particle near the bubble wall is sufficient to compensate for its rest mass energy at infinity. The na\"ive vacuum in the zero-charge sector is then unstable against spontaneous pair creation, and in the subsequent section we argue that the bubble discharge rate is unsuppressed.

We begin by parametrizing the 6D spacetime~(\ref{eq:dsfull}) as
\begin{align}
ds^2=G_{\mu\nu}dx^\mu dx^\nu+Vd\chi^2\;
\label{eq:fivemetric}
\end{align}
where $V=U(\rho)$ by comparison with~(\ref{eq:spatial}). The dimensional reduction of the three-form flux gives rise to an ordinary 5D electric field, and we choose a gauge where the field arises from a scalar potential vanishing at infinity,
\begin{align}
A_t=\frac{\sqrt{L}Q_0}{4\pi^2\rho^2}\;.
\label{eq:At}
\end{align}

Dimensionally reducing the worldsheet action for a wound string of tension $T$ with no $\chi$ excitations, the corresponding worldline action for the free particle is
\begin{align}
m \int d\tau \sqrt{V}\sqrt{-G_{\mu\nu}\partial_\tau x^\mu\partial_\tau x^\nu}
\end{align}
where $m=TL$ for $L\gg1/\sqrt T$. Therefore, in our toy model we introduce a charged scalar with action
\begin{align}
S[\Phi]=\int d^5x\sqrt{-G}\left(-G^{\mu\nu}(D_\mu\Phi)^*(D_\nu\Phi)-m^2 V |\Phi|^2\right),
\end{align}
where $D_\mu=\partial_\mu+i q A_\mu$ for a particle of charge $q$. Here  $q$ has mass dimension $-1/2$. We treat $\Phi$ as a probe, neglecting backreaction on the metric and gauge field.

The Klein-Gordon equation for $\Phi$ is 
\begin{align}
({\cal D}^2-m^2 V)\Phi 
     =(\Box+2iqA^t\partial_t-q^2 A^t A_t -m^2 V )\Phi=0\;,
     \label{eq:KGfull}
\end{align}
and the energy is 
\begin{align}
E&=\int d^4 x \sqrt{-G}\left[-G^{tt} |\partial_t\Phi|^2+G^{ii} |\partial_i\Phi|^2+\Phi^*\left(q^2G^{tt}A_t^2 +m^2 V\right)\Phi\right]\;.
\label{eq:energy}
\end{align}
The kinetic, mass, and gradient terms in the energy are  positive, although the mass term is suppressed by a factor of $V$ near the bubble wall. The $A_t^2$ term, which arises from the last term in the gauge-invariant charge density $J_t=i \Phi^* \partial_t\Phi-i \partial_t\Phi^* \Phi-2 qA_t|\Phi|^2$, is negative, indicating that small perturbations can lower the energy if this term dominates. Note, however, that the ``potential energy operator" $-\nabla^2+q^2G^{tt}A_t^2 +m^2 V$ differs from the fluctuation operator appearing in the equation of motion by a term $ -2iqA^t\partial_t$. Thus negative energy perturbations do not immediately imply complex frequencies or the exponentially growing modes characteristic of classical instabilities. 

We will look for $s$-wave solutions to~Eq.~(\ref{eq:KGfull}). Setting $\Phi=\phi(\rho) e^{i\omega t}$ we obtain
\begin{align}
h U \phi '' + \left(U
   h'+\frac{h U'}{2}+\frac{3 h U}{\rho }\right)\phi '+ \left(\frac{\left(q A_t+\omega \right){}^2}{h}-m^2 U\right)\phi=0\;.
\label{KGeq}
\end{align}
This equation admits bound states of finite Klein-Gordon norm (charge). 
The energy and charge of a bound mode is
\begin{align}
E_\phi&=\int d^4 x \sqrt{-G}G^{tt}\phi^*\left(-2\omega^2-2qA_t\omega\right)\phi\nonumber\\
\calQ_\phi&=-q\int d^4 x \sqrt{-G}G^{tt}J_t=-\frac{q}{\omega} E_\phi\;.
\label{eq:charge}
\end{align}
Here we have used the equation of motion and assumed Dirichlet conditions at $\rho=\rho_0$, which is sufficient to prevent energy or charge flux through the bubble wall and simplifies the analysis of modes. Below we will relax this condition and allow charged matter to fall into the bubble,  annihilating some of the bubble charge.

Now we can study the single-particle ground states in the charge $\pm$ sectors as a function of the charge to mass ratio, 
\begin{align}
w\equiv q M_5^{3/2}/m.
\end{align} 
Here $M_5$ is the 5d Planck scale, $M_5^3=LM_6^4$. Before proceeding, we note that the analysis of Eq.~(\ref{KGeq}) in the static background~(\ref{eq:spatial}),(\ref{eq:dsfull}) is only meaningful for the following  hierarchies of mass scales:
\begin{align}
M\gg M_5\gg m\gg\sqrt{m/L} \gg Q^{-1/2}\gg L^{-1}\;.
\label{eq:hierarchy}
\end{align}
The first and second inequalities allow us to treat the bubble as a fixed classical background on which the single particle states are a perturbation. The third inequality allows us to set the wound string mass to $m=TL\gg\sqrt{T}$, and the fourth imposes the requirement that the spacetime curvature near the bubble wall is below the string scale $\sqrt{T}$. The final inequality arises from Eq.~(\ref{eq:stab}) and the requirement that tunneling transitions to larger bubbles are suppressed (cf. Fig.~\ref{fig:Mrho0}). Subsequently we will mostly work in 5d Planck units, $M_5=1$.

\begin{figure}[t!]
\begin{center}
\includegraphics[width=0.6\linewidth]{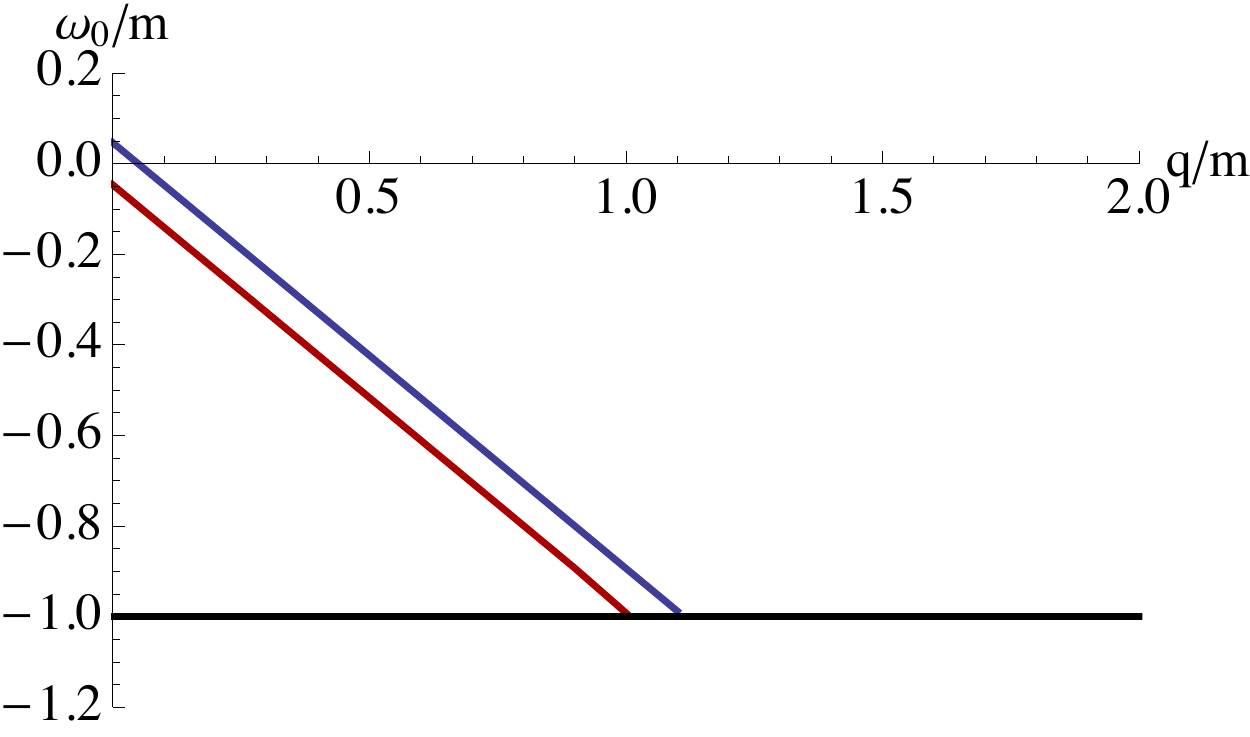}
\caption{Ground state frequencies  obtained by numerical solution of Eq.~(\ref{KGeq}). Normalized to $m$, the frequencies depend only on the combinations $q/m$, $Q/Q_{max}$, and $mL$. We take a point where $mL=100$ and $Q/Q_{max}\approx 1/4$, scanning over $q/m$ in $5d$ Planck units. The upper solid line (blue) shows $\omega_0$, the ground state frequency (energy) in the charge $-q$ sector. The lower solid line (red) shows the ground state frequency (-energy) in the charge $+q$ sector. The dashed line denotes the charge $+q$ continuum. For $q=0$, $\omega_0\sim \sqrt{T}\ll m$. For small positive $q$, $\omega_0$ crosses zero. At some $q/m\sim\calO(1)$, $\omega_0$ crosses into the positive-charge continuum. At this point the energy of an oppositely-charged pair vanishes, and the system becomes unstable against spontaneous pair creation.
} 
\label{fig:numerics}
\end{center}
\end{figure} 

Since the metric functions~(\ref{eq:U}),~(\ref{eq:h}) are somewhat involved and the exact solution for $\rho_0$ is the root of a cubic, it is simplest to perform  detailed analysis numerically.  Normalized to $m$, the eigenfrequencies depend only on the dimensionless ratios $w$, $Q/Q_{max}\ll1$, and $mL\gg 1$.
For typical sets of parameters we can solve~(\ref{KGeq}) numerically for the bound states. In Fig.~\ref{fig:numerics} we show the frequencies of the charge $\pm q$ ground states as a function of $q$. We see that in this example, $\omega\ll m$ for $q=0$ and decreases approximately linearly as $q$ is increased.

These properties are straightforward to understand physically. For $q=0$ and vanishing energy flux through the bubble wall, the $s-$wave spectrum includes a discrete set of normalizable bound modes of $|\pm\omega|<m$. The lowest modes have $|\omega|\sim \sqrt{T}$, reflecting the fact that wound strings near the cigar tip are close to becoming unwound (and indeed would become unwound with any asymmetric perturbation). In the hierarchy~(\ref{eq:hierarchy}), these modes are deeply bound, $|\omega|\ll m$. For small positive $q$, the spectrum shifts downward. Positive frequencies, corresponding to negative charges, become more tightly bound, while their negative frequency counterparts shift toward $\omega=-m$. Since the modes are deeply bound, we can approximate the electrostatic potential energy term in the Klein-Gordon equation by its value at the bubble wall, $qA_t(\rho_0)$. Then the bound mode frequencies decrease as
\begin{align}
\frac{d\omega}{dq}\approx -\frac{Q}{2\rho_0^2}\approx -1.
\end{align}

Denoting the lowest mode in the negative charge sector by $\omega_0$, for some small value of $q$, the ground state energy $\omega_0\rightarrow 0$. At this point the negative charge state has binding energy that completely cancels its asymptotic rest mass. However, because of charge conservation, this is not yet enough to indicate an instability in the charge-zero sector. As $q$ is increased further the ground state energy drops below zero. Once $q\sim m$, we find $\omega_0\rightarrow -m$. At this point the bound state now has sufficiently negative energy to compensate for the rest mass of an additional positive charge at infinity.

For larger $q$, $\omega_0$ cannot simply fall below $-m$. These solutions are unbound and correspond to positive-charge scattering states. This situation is  identical to the physics of high-$Z$ ``over-critical" nuclei~\cite{Rafelski:1976ts}. It is energetically favorable to spontaneously produce opposite-charge pairs, and the na\"ive zero-particle ground state is unstable.

\subsection{Discharge Rate}

Heuristically, the discharge process can be thought of as pair creation near the bubble wall. The negative charge annihilates some of the bubble charge, while the positive charge tunnels out and escapes to infinity. The whole process must conserve energy, so the energy $\epsilon_+$ of the escaping positive charge must be compensated by the reduction in mass of the bubble,
\begin{align}
\epsilon_+=-\Delta M =  \frac{qQ}{2\rho_0^2}
\end{align}
assuming $q\ll Q$ in 5D Planck units.  ($\Delta M$ is also equal to the classical energy of a negative charge at rest at the bubble wall, but we do not need this interpretation.) 

If the escaping ``positron"  encounters a potential barrier, then the discharge rate is  proportional to a  tunneling exponent which can be straightforwardly estimated in the WKB approximation.

The worldline action for the positively-charged state moving in the radial direction $\rho(t)$ is
\begin{align}
S=\int dt\left(-m\sqrt{V}\sqrt{-G_{tt}-G_{\rho\rho}\dot\rho^2}-\frac{qQ}{2\rho^2}\right)\;.
\end{align}
The Hamiltonian is
\begin{align}
H&=\frac{-m\sqrt{V}G_{tt}}{\sqrt{-G_{tt}-G_{\rho\rho}\dot\rho^2}}+\frac{qQ}{2\rho^2}\nonumber\\
&=\sqrt{-G_{tt}(G^{\rho\rho}\pi^2+m^2V)}+\frac{qQ}{2\rho^2}
\end{align}
where in the second line we have written the Hamiltonian in terms of the canonical momentum $\pi(t)$.  In the WKB approximation, the tunneling amplitude through classically forbidden regions is proportional to 
\begin{align}
\Gamma\sim\exp{(i\int_{\rho_-}^{\rho_+} \pi d\rho )}
\label{eq:wkbexp}
\end{align}
where 
\begin{align}
\pi=\sqrt{-G_{\rho\rho}\left(G^{tt}\left(\epsilon_+-\frac{qQ}{2\rho^2}\right)^2-m^2V\right)},
\label{eq:wkbpi}
\end{align}
and $\rho_\pm$  are the classical turning points.  In classically forbidden regions the integral is complex, and the factor~(\ref{eq:wkbexp})  suppresses the tunneling rate. 

The turning points are located at $\rho_-=0$ and 
\begin{align}
\rho_+=\frac{Q\rho_0\sqrt{q^2-m^2}}{\sqrt{q^2Q^2-4m^2\rho_0^4}}
\end{align}
Since $Q< 2\rho_0^2$, the outer turning point is only finite (and therefore the pair production rate is only nonzero) if
\begin{align}
q\gtrsim m \frac{2\rho_0^2}{Q}\approx m\;.
\end{align}
If this inequality is satisfied, the WKB integral gives
\begin{align}
\int_0^{\rho_+} \pi d\rho \approx i\sqrt{\rho_0^2-\frac{Q^2}{4\rho_0^2}}\left(-m+\frac{qQ}{2\rho_0^2} {\rm tanh}^{-1} \left(\frac{2\rho_0^2}{q Q}m\right)\right)\;.
\label{eq:intwkbexp}
\end{align}

Let us first examine this exponent in the limit $Q/L^2\ll 1$. This is the limit in which the bubble is most stable against tunneling to a larger, expanding bubble. Expanding~(\ref{eq:intwkbexp}) in $Q/L^2$, we find
\begin{align}
\Gamma\sim \exp\left({-\frac{\pi m Q (-m+q~{\rm tanh}^{-1}(m/q)}{m L}}\right)\sim \exp\left[{-\pi \left(\frac{Q}{L^2}\right) \left(\frac{m^2}{q^2}\right)(mL)}\right]
\label{eq:wkbrate}
\end{align}
where in the last step we have approximated $q/m\gtrsim{\rm~few}$. 

For comparison, in the small $Q$ limit the decay rate into a larger, expanding bubble should be well-approximated by the decay of the ordinary KK vacuum into neutral bubbles. (For larger $Q$, the rate will be faster.) The rate for this process is of order 
\begin{align}
\Gamma_{\rm Witten}\sim e^{-M_6^4 L^4}\sim e^{-M_5^3 L^3}. 
\end{align}
For $q/m\gtrsim 1$ and recalling the hierarchies in Eq.~(\ref{eq:hierarchy}), the discharge rate~(\ref{eq:wkbrate}) satisfies
\begin{align}
\Gamma> e^{-mL}\gg \Gamma_{\rm Witten}\;.
\end{align}

We cannot take $Q$ smaller than $L/m$ if we want to keep curvatures everywhere below the string scale. In this limit, the discharge rate becomes
\begin{align}
\Gamma\sim \exp{\left(-\frac{\pi m^2}{q^2}\right)}\;.
\end{align}
In other words, when the  decay rate into expanding bubbles is minimized, the discharge process is unsuppressed, if the WGC is satisfied.

\begin{figure}[t!]
\begin{center}
\includegraphics[width=0.5\linewidth]{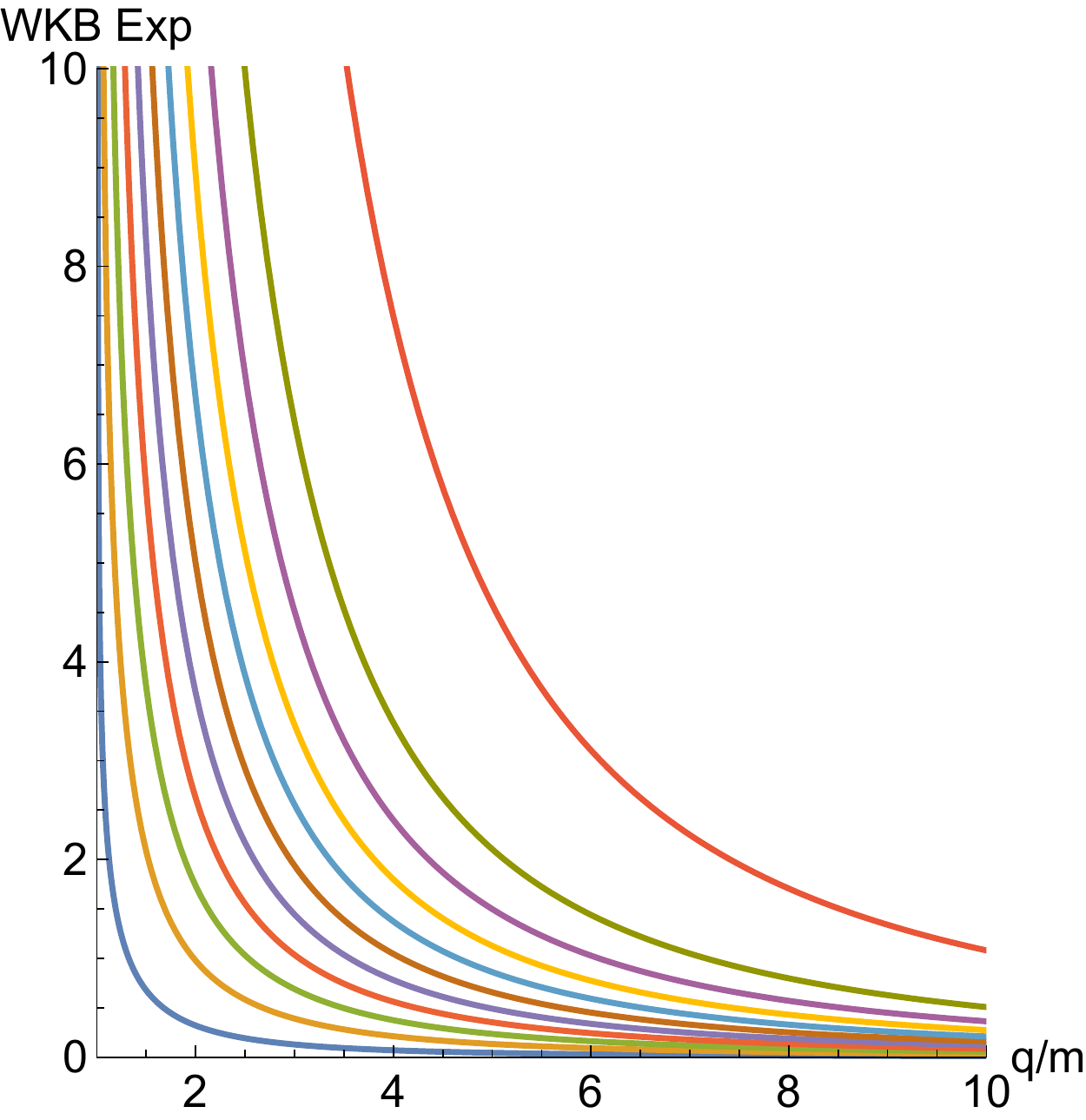}
\caption{ The WKB exponent in Eq.~(\ref{eq:intwkbexp}) for $mL=10^3$ as a function of $q/m$. From bottom to top the curves correspond to $Q=(L/m,\dots,Q_{max})$. Higher curves (larger $Q$) have increasingly suppressed discharge rates, but are increasingly unstable against tunneling to expanding bubbles. (The top curve is completely unstable.) We see that when the WGC is satisfied by a modest amount the exponent is generally small and the discharge rate is fast. 
} 
\label{fig:WKBexp}
\end{center}
\end{figure}

We can also assess the WKB exponent numerically for other values of the parameters. In Fig.~\ref{fig:WKBexp} we show the exponent for $L/m<Q<Q_{max}$ and $mL=10^3$ as a function of $q/m$. The exponent is typically not large if $q/m\gtrsim{\rm few}$. 

Determining the endpoint of the discharge process requires consideration of backreaction effects. Qualitatively, we expect the bubble to collapse into a black string. It is remarkable that in the limit where one instability is made small, a new one appears with large rate.

\section{Charged Dilatonic Black Holes}
The KK bubbles considered above sample an infinite distance in moduli space. We can also consider charged dilatonic black hole spacetimes in which the dilaton excursion is finite but can be made arbitrarily large~\cite{Garfinkle:1990qj}.

The electrically charged solutions of~\cite{Garfinkle:1990qj} are given by the dilaton, electric field, and Einstein frame metric:
\begin{align}
\phi=-\phi_0+\frac{1}{2}\log\left(1-\frac{Q^2 e^{-2\phi_0}}{Mr}\right)\;,
\end{align}
\begin{align}
F_{tr}=E=\frac{e^{-2\phi_0}Q}{r^2}\;,
\end{align}
\begin{align}
g_{\mu\nu}dx^\mu dx^\nu=-f\, dt^2 + f^{-1}\, dr^2+r\left(r-\frac{Q^2 e^{-2\phi_0}}{M}\right)(d\theta^2+\sin^2\theta d\phi^2)\;,
\end{align}
where $f=1-2M/r$ and the dilaton Lagrangian is $2(\nabla \phi)^2+e^{-2\phi}F^2$. The extremal limit is $Q_{ext}=\sqrt{2}e^{\phi_0}M$ and there is a horizon at $r_+=2M$. Let us define an extremality parameter $y=Q/Q_{ext}$. In terms of $y$ the dilaton excursion between infinity and $r_+$ is
\begin{align}
\Delta\phi=\frac{1}{2}\log(1-y^2)\;.
\end{align}
We see that $|\Delta\phi|$ is finite and large for near-extremal black holes, and $\Delta\phi\rightarrow-\infty$ in the extremal limit. The Maxwell term in this theory is $e^{-2\phi}F^2$, and it is convenient to canonically normalize it at infinity. We define $\hat F=e^{\phi_0}F$ and $\hat Q=e^{-\phi_0}Q$ such that $\hat E=\hat Q/r^2$ and the extremal limit is $\hat Q_{ext}=\sqrt{2}M$. The Hawking temperature is
\begin{align}
T=\frac{1}{8\pi M}\;.
\end{align}
Despite the finite temperature, the Kretschmann scalar behaves as $M^{-4} e^{-4\Delta\phi}$ near the horizon for $|\Delta\phi|\gg 1$. Subplanckian curvatures thus imply $|\Delta\phi|\lesssim\log(M)$.

Now we introduce a charged particle of mass $m$ and charge $q$ coupled to the gauge field $\hat A_t=\hat Q/r$. 
The black hole may be hot ($m/T\lesssim 1$) or cold ($m/T\gtrsim 1$). For simplicity we assume a minimally coupled charged scalar field, which is meaningful if $m\ll 1$.

In the hot case, the emission probability for charges $\pm q$ is proportional to a Boltzmann factor of the form
\begin{align}
e^{-\frac{1}{T}\left(m\pm\frac{qQ}{r_+}\right)}= e^{-\frac{m}{T}\left(1\pm\frac{q}{\sqrt{2}m}(1-e^{2\Delta\phi})\right)}.
\end{align}
If $|\Delta\phi|\gg 1$, the discharge rate is fast if the WGC is satisfied and the black hole is hot.

Now we consider the case $m/T\gg 1$. Here the discharge rate is governed by the Schwinger process, and the rate exponent can be determined by barrier penetration arguments for a mode of frequency equal to the electrostatic potential energy at the horizon, $\omega_+=-qQ/r_+$~\cite{Gibbons:1975kk}. For $\omega_+$ to be a scattering state, we must have $q>\sqrt{2}m$ near extremality. 

The Klein-Gordon equation for the $s$-wave mode of frequency $\omega_+$ is
\begin{align}
\Phi''(r)+W\Phi(r)=0\;,\;\;\;\;\;W=\frac{q^2}{2}-\frac{m^2r}{r-2M}+\frac{(Me^{2\Delta\phi})^2}{(r-2M)^2(r-2M+2Me^{2\Delta\phi})^2}\;.
\end{align}
Here we have put the equation in normal form and taken $e^{2\Delta\phi}\ll 1$.
The barrier $W<0$ extends approximately from $r\sim 2M+e^{2\Delta\phi}M \equiv \alpha$ to $r\sim 2M\left(1+ \frac{2m^2}{q^2-2m^2}\right)\equiv \beta$.   In the WKB approximation, the barrier penetration factor is 
\begin{align}
e^{-2\int_\alpha^\beta \sqrt{W}dr}.
\end{align}
We can approximately evaluate the WKB integral by splitting it into regions where the last two terms in $W$ dominate ($\equiv W_{23}$, valid near $\alpha$) and where the first two terms in $W$ dominate ($\equiv W_{12}$, valid near $\beta$). The two regions overlap where the first and third terms are of similar order, near $r\sim 2M+1/\sqrt{2}q$. Putting the pieces together and keeping only the dominant terms, we find the production rate is of order
\begin{align}
\Gamma_{Einstein} \sim e^{-2\int_\alpha^\gamma \sqrt{W_{23}}dr-2\int_\gamma^\beta \sqrt{W_{12}}dr}\approx e^{-\frac{2\sqrt{2}\pi  M m^2}{\sqrt{q^2-2m^2}}}\;.
\label{eq:einsteinrate}
\end{align}
This is similar to the Schwinger exponent $\pi m^2/q\hat E$ arising from a constant field of magnitude $\hat E=\hat Q/r_+^2$. For a large, cold black hole, the rate is exponentially small for $q\sim m$.

We conclude that a large field excursion $|\Delta\phi|\gg1$ requires a large source. Exponentially large sources are required to control curvature invariants at the horizon. In the presence of a light particle  of mass $m$ satisfying the weak gravity conjecture, it is also possible for the black hole to rapidly discharge. Combining the requirements of slow discharge and low curvature, we obtain the bound
\begin{align}
M\gg {\rm max}(e^{|\Delta\phi|},1/m).
\end{align}
This is reminiscent of other indications that large localized field excursions can be sustained around exponentially large sources~\cite{nicolis}.

Other similar dilatonic black hole solutions can be obtained, including different dilaton couplings $e^{-2a\phi}F^2$ and general dyonic charges~\cite{Ivashchuk:1999jd,Abishev:2015pqa} (see also~\cite{Loges:2019jzs} for a recent analysis in the context of the WGC).\footnote{We thank Gary Shiu for bringing these solutions to our attention.} In some cases simple analytic solutions are known. In the magnetic case, similar results are obtained. In the dyonic case, it is possible to have a finite dilaton excursion in the extremal limit. However, the curvature at the horizon is still controlled by the mass of the black hole and the amplitude of the dilaton excursion, in such a way that analogous bounds of the form $|\Delta\phi|<\log(M)$ still hold.

\section{Discussion}
Both neutral and charged KK bubbles sample infinite distances in moduli space in finite spatial regions with size $R$ of order the bubble radius. Neutral bubbles are classically unstable, and we have argued that charged bubbles are destabilized in the presence of charged matter with $q/m\gtrsim 1$. Dilatonic black holes in a controlled EFT have a finite excursion $ |\Delta\phi|\lesssim |\log(M_{BH})|$.

We have not discussed KK monopoles~\cite{sorkin,GP}, but they provide another interesting example. They are stable and sample an infinite distance in moduli space, but only at a single point. The KK scalar modulus diverges as 
\begin{align}
|\Delta \phi| \sim |\log(r/L)|
\end{align}
near the center of the monopole. Access to distance scales shorter than $L$ is required to see the excursion, but in principle this is permissable since we do not insist on dimensional reduction. The accessible excursion is ultimately limited by the short-distance cutoff on the semiclassical KK theory, $|\Delta \phi| \lesssim |\log(\Lambda L)|$.

Motivated by these examples, we conclude with a loose conjecture: quantum gravity in asymptotically flat space requires a general bound on localized, (meta)stable moduli space excursions of the form 
\begin{align}
|\Delta\phi|\lesssim |\log(R\Lambda)|
\end{align}
where $\phi$ parametrizes the modulus, $R$ is a scale characterizing the minimal region enclosing the excitation, and $\Lambda^{-1}$ is a short-distance cutoff on local quantum field theory. In a sense, the swampland distance conjecture applies to the limit $R\rightarrow\infty$, where the moduli are moved everywhere in space, and the bound is trivially satisfied.  This bound is consistent with the Newtonian analysis of~\cite{nicolis}, but we have seen that it is less trivial in general relativity, and requires the WGC in some cases. 

There also appears to be another connection with the SDC: large excursions are typically confined near surfaces or points, rather than being spread over finite bubble volumes. Consequently, access to the excursion requires  access to short distance scales. In the KK examples, this also implies access to a  tower of KK states scaling exponentially with the observable excursion.

\vskip 1cm
\noindent
{\bf Acknowledgements:}  PD thanks B. Heidenreich, G. Horowitz, M. Montero, M. Reece, G. Shiu, and I. Valenzuela for useful discussions. This work was supported by NSF grant PHY-1719642.
\bibliography{dipole_refs}{}
\bibliographystyle{utphys}

\end{document}